\documentclass[
    ,final            
  ]
  {aipproc}

\layoutstyle{6x9}

\def\x{{\bf x}}
\def\y{{\bf y}}
\def\D{{\bf D}}
\def\A{{\bf A}}
\def\E{{\bf E}}
\def\bdel{{\mbox{\boldmath $\nabla$}}}

\begin{document}

\title{Aspects of Confinement: a Brief Review}

\author{Eric S. Swanson}{
  address={Department of Physics and Astronomy, University of Pittsburgh, Pittsburgh PA, 15260}
}

\begin{abstract} 
A brief and biased overview of the phenomenon of confinement in QCD is presented in three
parts: (1) the definition of confinement, (2) properties of confinement, (3) ideas of confinement.
The second part chiefly consists of a brief review of recent lattice computations related to 
confinement while the third summarizes some of the current analytical approaches to understanding
confinement. These include the Dyson-Schwinger formalism in Landau gauge, Hamiltonian 
QCD in Coulomb gauge, and the vortex picture of confinement.
\end{abstract}

\maketitle

\section{Definitions of Confinement} 

Confinement is the poster boy of nonperturbative physics for good reason: 
it is associated with a linear potential with a string tension,

\[
\sigma \propto \Lambda^2 {\rm e}^{-\int {dg \over \beta(g)}}
\]
which is nonperturbative in the coupling.

Although this statement has intuitive appeal, one must be careful in defining
confinement. For example, it is often loosely  defined as the absence of 
free quarks in nature. But it is conceivable, and even possible, that there exists
a coloured scalar particle which can form bound states with quarks. The 
resultant particles would then  carry flavour and fractional electric charge\cite{G}, which 
is likely not the intent of the definition. 

Similarly, requiring that all observable particles be colour singlets  encompasses
gauge theories in both the confinement and the Higgs phases. The latter also manifests
colour singlet states because colour charge is completely screened. 

One may hope to improve the situation by 
focussing on the expected physical properties of confinement. For example, the
appearance of a long range linear potential between quarks is a reasonable requirement.
Of course, the problem here is that string breaking will occur once the potential
energy approaches the quark pair creation threshold. The conventional way out of this
difficulty is to consider the work to separate two quarks as the quark masses approach
infinity. Thus one is led to the strange position of defining quark confinement in a 
limit which removes quarks from the theory.   

\subsection{Centre Vortices}

It is traditional to implement the last definition in lattice gauge theory with the aid of
the Wilson loop:

\begin{equation}
\langle WL\rangle = \int DU_\ell {\rm tr} [ \prod_{i\in { C}} U_i]{\rm e}^{-S_{YM}}
\end{equation}
where $C$ is a large planar loop and the pure gauge action is given by

\begin{equation}
S_{YM} = {6\over g^2} \sum_{P} \left( 1 - {1\over N_c}\Re {\rm tr}[ U_P ]\right).
\end{equation}
Here the gluonic degrees of freedom are represented by link variables $U_\ell = 
\exp(ig T^a A_\mu^a(x))$ where $\ell$ represents a link on a spacetime lattice which starts
at the point $x$ and points in the $\hat \mu$ direction and the lattice spacing has been set to
unity.  The sum in the action is
over plaquettes, $P$, which are the smallest closed loops permitted on the lattice. $U_P$
is a product of link variables around a plaquette and is thus a lattice implementation of the two-forms of gauge theories. It is measurements of the Wilson loop which have
provided the most compelling demonstrations of the confinement phenomenon.

\begin{figure}[h]
\includegraphics[height= 7 true cm]{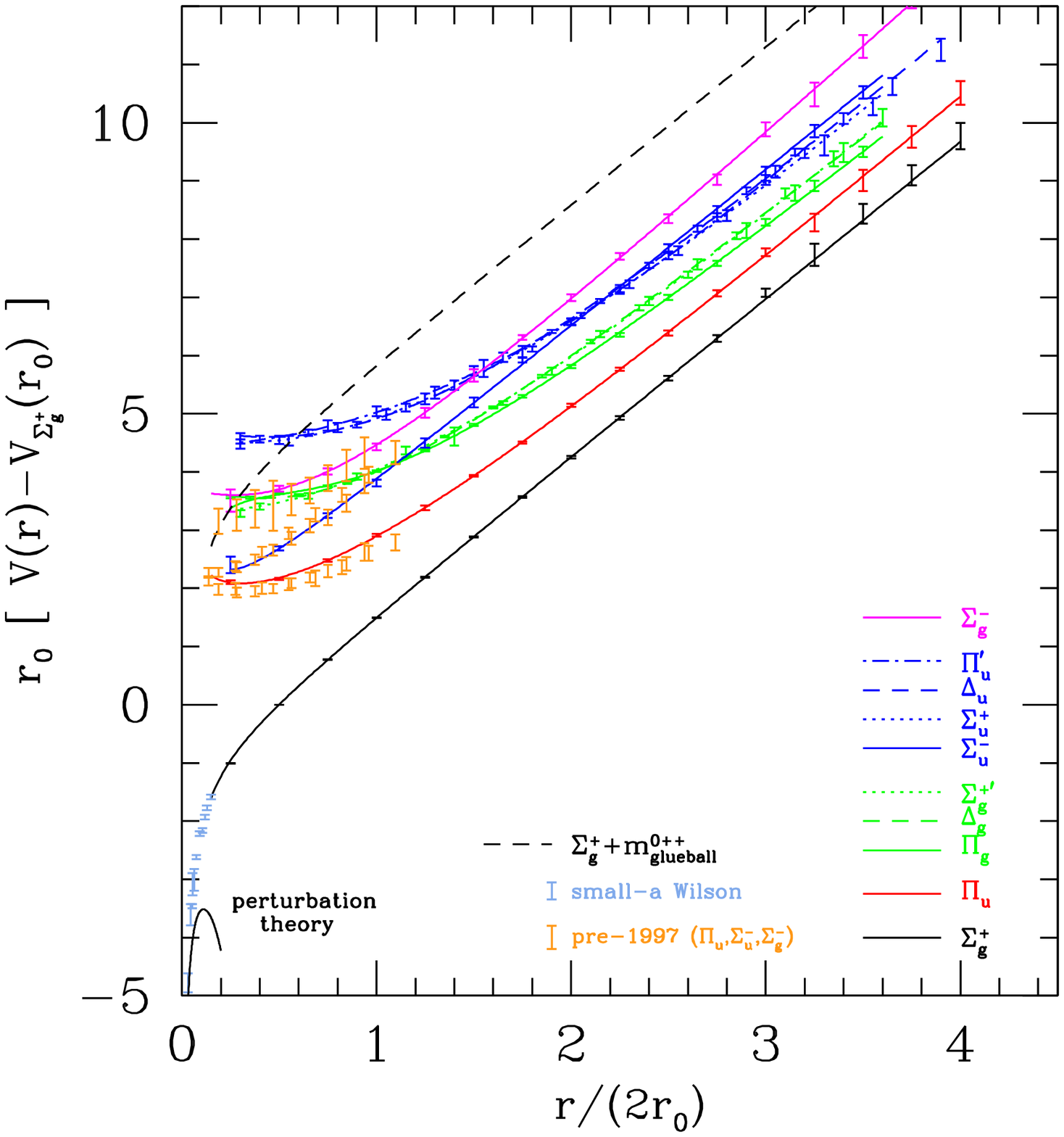}
\includegraphics[height= 7 true cm]{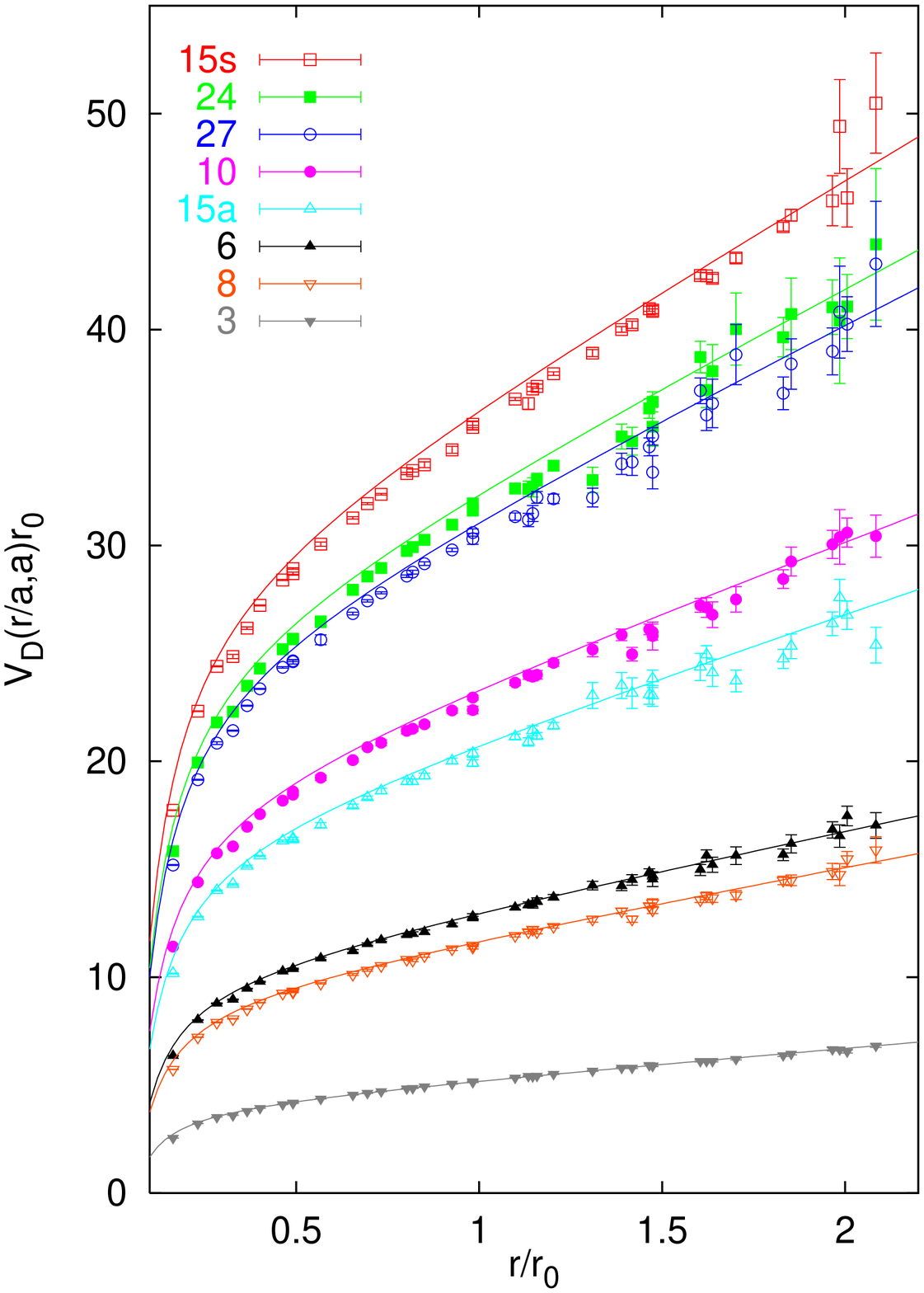}
\caption{(left) Wilson Loop Measurements of Various Static Quark Potentials\protect\cite{JKM}. $\ \ $
(right) Casimir Scaling of Confinement\protect\cite{Bali2}.}
\label{VRplot}
\end{figure}

The lattice definition of QCD and confinement makes it especially easy to see the
significance of the centre group of SU(3). The centre is defined as the set of all
elements of the gauge group which commute with all other elements. Evidently these
are given by the set $ \{ z = \exp(2\pi n / N_c) {\cal I}\}$ where $n = 0,\ldots,N_c-1$. This set
forms the 
group $Z_{N_c}$. It is clear that multiplying all temporal link variables on a particular
time slice  by an element of the group , $U_t(t_0,\x) \to z U_t(t_0,\x)$ does not change the
value of a plaquette, implying that QCD is
invariant under global $Z_{N_c}$ transformations. 

We now consider a Wilson loop which extends in the temporal
direction for the entire length of the lattice (and, because of the imposition of
boundary conditions, encircles the lattice), called a Polyakov line. If one makes the transformation discussed
above it is clear that the Polyakov line picks up a phase:
$PL(\x) \to z PL(\x)$. Since the theory is invariant under such a transformation one
must have either $\langle PL(\x)\rangle = 0$ 
if the vacuum of QCD shares the symmetry of the action
or $\langle PL(\x)\rangle \neq 0$
if the theory is in the $Z_{N_c}$ broken phase.

The connection with confinement arises once it is realized that the Polyakov line 
measures the free energy of a static quark at the position $\x$: $\langle PL(\x) \rangle = 
\exp(-E T)$ where $T$ is the temporal extent of the lattice. But the free energy of an isolated confined quark is infinite, thus {\it QCD must be in the unbroken $Z_{N_c}$ symmetry phase
if it is confining}.

It is useful to note that $Z_{N_c}$ transformations are an example of ``singular gauge
transformations'' (which are {\it not} gauge transformations!), namely they may be
generated by performing a gauge transformation which is periodic modulo a $Z_{N_c}$ 
phase factor:

\begin{equation}
U_t(t,\x) \to g(t,\x) U_t(t,\x)g^\dagger(t+1,\x)
\end{equation}
with  $g(T+1,\x) = z^*g(1,\x)$. For example the symmetry transformation
$U_t(t_0,\x) \to z U_t(t_0,\x)$ may be achieved by setting $g(t,\x) = 1$ for $t\leq t_0$ and 
$g(t,\x) = z^*$ for $t > t_0$. Singular gauge transformations are of central interest to the
study of confinement because the structures associated with them are often postulated to cause
confinement. We shall return to this in section III below.

\subsection{Kugo-Ojima Criteria}

The final definition of confinement to be considered here was first investigated by Kugo and
Ojima shortly after the invention of QCD\cite{KO}. Their starting assumption was that only
BRST singlets may be allowed as physical states if confinement is to hold. A sufficient
condition for this is that the ghost propagator in Landau gauge is enhanced in the 
infrared:

\begin{equation}
D_G(k) = -{1\over k^2} {1 \over 1+ u(k^2)} \equiv -{G(k^2)\over k^2}
\end{equation}
with $u(0) = -1$. This may be related to the gluon propagator via Dyson-Schwinger equations 
and implies
that there is an infrared suppression of the 
transverse gluon correlator:
\begin{equation}
D_{\mu\nu}(k) = {Z(k^2)\over k^2} \left(\delta_{\mu\nu} - {k_\mu k_\nu \over k^2} \right)
\end{equation}
with $Z(k^2)/k^2 \to 0$ as $k\to 0$\cite{TK}. 

It is interesting to note that these criteria are consistent with results derived by
Zwanziger in noncovariant gauges\cite{Z}.

\section{Properties of Confinement} 

At present the only reliable method for determining properties of confinement is
with lattice computations.  Indeed, the detection of confinement (with the Wilson
loop) was one of the first lattice computations and it continues as the standard bearer
of lattice gauge theory.  

Recent investigations have revealed much beyond the static
quark interaction. Perhaps the most famous examples are plots of action or field density
which clearly show the formation of tubes of gluonic flux forming between the colour
source and sink\cite{Bali}. Of course, such tubes fit in very well with the naive notion of linear quark
confinement. More interesting is recent investigations of the {\it dynamics} of flux tubes,
as represented by the higher surfaces in Fig. \ref{VRplot}(left). These may be 
interpreted as adiabatic energy surfaces describing hybrid mesons.

Another type of flux tube investigation places small kinks in Wilson loops to study
the spin-dependence of the long range confining force\cite{WLspin}. The results are
consistent with the supposition that confinement is of a Lorentz scalar nature (it
is important to note, however, that mapping the lattice results to this type of
interaction is an effective, low energy approximation only).

\vskip .5 true cm
\noindent
{\it Casimir Scaling}

Figure \ref{VRplot}(right) shows in a compelling way the property of {\it Casimir scaling} of
confinement. The figure was obtained by measuring the Wilson loop for sources in various
representations of SU(3). The interaction between colour triplets is the lowest surface
in the figure and forms the template for the others. In the figure one sees higher 
surfaces with
sources in the 8, 6, 15$_A$, 10, 27, 24, and 15$_S$ representations. The curves are obtained
by multiplying a fit to the lowest (fundamental representation) surface by the quadratic
Casimir, ${\cal C}_R^2 = \langle R |T^a T^a|R\rangle$ divided by ${\cal C}_F^2$. The quadratic Casimir is given by $(p^2+q^2+pq)/3 + p + q$
where $(p,q)$ is the Dynkin index of the representation.
The agreement is remarkable and is a strong indication that the colour structure of 
confinement may be written as

\[
\int \bar\psi T^a \psi \ldots \bar \psi T^a \psi
\]
where the ellipsis represents Lorentz and spatial dependence.

\vskip .5 true cm
\noindent
{\it String Behaviour}

In pre-QCD days strings were invoked as a fundamental theory of QCD. With the advent of
QCD, strings have survived as an effective description of flux tubes. In particular
one expects that 

(i) transverse flux tube profiles should be logarithmically divergent
with the tube length (due to string `roughening');

(ii) the ground state potential should exhibit a universal `Luescher term', $-\pi/(12 R)$;

(iii) the string excitation spectrum should be that of bosonic string modes, $\pi/R$.

\begin{figure}
\includegraphics[height= 6 true cm]{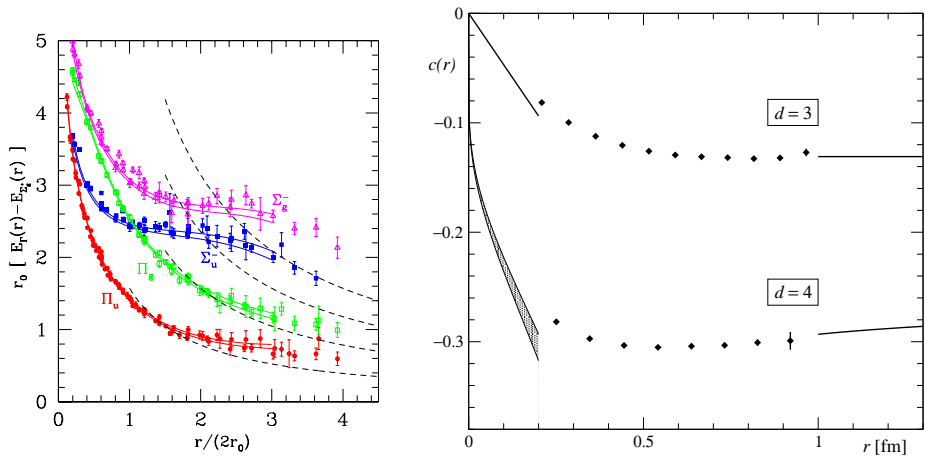}
\includegraphics[height= 6 true cm]{luescher.eps}
\caption{(left) Hybrid Surface Energy Differences. 2$r_0$ is roughly 1 fermi\protect\cite{JKMstring}. $\ \ $  (right) The Luescher Term in 4-d ($-\pi/12$) and 3-d ($-\pi/24$)\protect\cite{LW}.}
\label{strings}
\end{figure}

Juge, Kuti, and Morningstar have carried out a detailed analysis of the relationship of the hybrid surfaces of  Fig. \ref{VRplot}(left) to string excitations\cite{JKMstring}. They have found
that surface excitations only have $\pi/R$ splittings for very large source separation
(roughly 4 fermi or greater); see Figure \ref{strings}(left). Furthermore, there is a cross over region at about 1 fermi
where the surfaces move from a perturbative behaviour (characterized by the `gluelump' spectrum) to a string-like behaviour. 

Point (ii) has been examined with a new algorithm by Luescher and Weisz\cite{LW}; their
main results are shown in Figure \ref{strings}(right). One sees that the expected behaviour of
the confinement potential is achieved at a source separation of roughly 1/2 fm or less.
String-like behaviour at such short distances may be called precocious and is very 
difficult to understand in light of the conclusion of Juge, Kuti, and Morningstar.  It 
appears that the existence of a $1/r$ interaction with the expected Luescher
coefficient is not necessarily indicative of string dynamics.

\vskip .5 true cm
\noindent
{\it Baryonic Flux Tubes}

\begin{figure}[h]
\includegraphics[height=4 true cm]{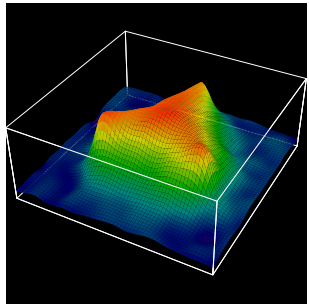}
\caption{The Baryonic Flux Tube. The distance between the tube junction and a quark
is approximately 1/2 fm\protect\cite{BFT}.}
\label{baryonFT}
\end{figure}

Finally, investigations of the static baryon interaction have begun. The chief
point of interest is whether the expected flux tubes form into a `Y' shape or a 
`$\Delta$' shape (ie.,  is the effective baryonic quark interaction two-body or
three-body?). This may be addressed by carefully examining the baryonic energy under
a variety of quark configurations. Figure \ref{baryonFT} shows the flux tube which 
arises in one such configuration (which seems to be an interpolation of Y and $\Delta$).
Current results are mixed, with some groups claiming support for the two-body hypothesis
\cite{Alexandrou} and some for the three-body hypothesis\cite{Suganuma}.  Finally, a strong
operator dependence in the flux tube profiles has been observed\cite{OW}, which clearly 
needs to be settled before definitive conclusions can be reached.

\section{Ideas of Confinement} 

The final portion of this report deals with analytic attempts to describe confinement.
Unfortunately, there is a great deal of prejudice concerning this issue, with
many people believing that any such attempts cannot succeed. This is not a useful point
of view to take and is, in fact, at odds with generations of successes in field theory, many-body physics, and condensed matter physics:

\begin{center}
{\it nonperturbative does not mean intractable!}
\end{center}

A trivial example is provided by elementary calculus. As is well known the function
$f(x) = {\rm e}^{-{1\over x}}$ has a zero radius of convergence and hence no sensible
`perturbation theory'. However, if a sufficiently clever theorist were
able to derive the `Dyson-Schwinger' differential equation
$f'(x) = {1 \over x^2} f(x)$, almost anyone could then derive any desired property of $f$.

More substantial examples are provided by the Gell-Mann -- Brueckner resolution to the
infrared divergence problem in the degenerate electron gas, the Galitskii method for dealing
with strong short range repulsive interactions in nuclear physics, and Laughlin's explanation
of the fractional quantum Hall effect. 

The point is this: QCD is described by  a Lagrangian and therefore all of the properties of
QCD are carried by its diagrams. To make progress nonperturbatively one must find an
infinite set of diagrams to `sum' which capture the majority of the physics one is interested 
in. It would be a completely new occurrence in physics if this approach should not work for confinement, requiring the introduction of a concept such as `intrinsic incomputability' (with
commensurate implications for attempts at deriving the `theory of everything').

In the following a very
brief summary of Dyson-Schwinger and Green's function approaches to confinement are 
presented. I shall also recall some salient features of the vortex picture of confinement. 
Space constraints forbid discussing many other popular ideas of confinement such as
the dual superconductor picture, merons, and monopoles.

\subsection{The Dyson-Schwinger Formalism}

There has been much progress in the Dyson-Schwinger approach to confinement in recent years.
The idea is to truncate the infinite series of Dyson-Schwinger equations (which are equivalent to the field theory) in such a way that the leading infrared behaviour of the ghost and
gluon propagators  may be reliably extracted. This endeavour is assisted by appealing to known
properties of nonperturbative vertices, either through Slavnov-Taylor identities, the application of BRST symmetries, or properties of special gauges (in particular, the ghost-gluon
vertex is not renormalized in Landau gauge). It has been found that including the ghost
propagator in the coupled set of D-S equations is important. One may then solve for the
infrared behaviour of the ghost and gluon propagators under various approximations or solve
the coupled integral equations numerically\cite{DS}. 

A convincing demonstration of confinement results. Analytic work finds  that the infrared propagators behave as

\begin{equation}
Z(k) \to \left({k^2 \over \sigma}\right)^{2\kappa}, \ k\to 0
\label{z}
\end{equation}
and
\begin{equation}
G(k) \to \left({\sigma\over k^2}\right)^\kappa, \ k \to 0.
\label{g}
\end{equation}
with $\kappa \approx 0.6$. This value is very stable under modification of the truncations
or Ans\"atze. Equations \ref{z} and \ref{g} demonstrate that the Kugo-Ojima confinement criteria are met. Furthermore, numerical solutions are in remarkable agreement with lattice
results\cite{AF}, indicating the utility of the truncations made.

\subsection{Green's Function Approach in Coulomb Gauge}

It has been 25 years since Gribov first noted that the imposition of Coulomb gauge in non-Abelian theories is beset with ambiguities\cite{gribov}. The ambiguity arises because
more than one solution to the gauge constraint, $\bdel\cdot \A = 0$ may exist. Gribov showed
that the condition for this to occur was that nontrivial solutions to the equation
$\bdel\cdot\D=0$ had to exist. Here $\D$ is the covariant derivative in adjoint 
representation.
Gribov proposed to resolve the ambiguity by restricting the gauge configurations to those with a
positive definite value of $\det(\bdel\cdot\D)$, called the Gribov region. He also noted that 
such a constraint
imposes a boundary in field space, and this boundary can affect the gluon dispersion relation,
causing confinement. 

Much progress has occurred since Gribov framed his conjectures. It is now known that the 
Gribov region does not resolve the Gribov ambiguity, rather a smaller region called the
fundamental modular region (FMR) is required. Furthermore, the FMR is contained in the
Gribov region, the FMR incloses the origin, it is convex, and its boundary sometimes 
coincides with the boundary of the Gribov region\cite{PvB}. Topological field
configurations are realized through the imposition of nontrivial boundary conditions at
the boundary of the FMR. Finally, Zwanziger has argued that physical matrix elements draw
their support solely from the intersection of the Gribov and fundamental modular boundaries\cite{Z4}. 

It is suggestive that the operator $\bdel\cdot\D$  appears in the instantaneous portion of the 
interaction in the 
QCD Hamiltonian in Coulomb gauge\cite{CL}:

\begin{equation}
V_{Coul} = -{1\over 2} \int d^3xd^3y\, {\cal J}^{-1}\rho^a(\x) \langle \x a | {g \over \bdel\cdot \D} \bdel^2 {g \over \bdel\cdot \D}|\y b\rangle {\cal J} \rho^b(\y)
\label{coul}
\end{equation}
where ${\cal J} = \det(\bdel\cdot \D)$ is the Faddeev-Popov determinant and the colour density is given by $\rho^a = -f^{abc}\A^b(\x)\cdot \E^c(\x)$. If vacuum field configurations are dominated
by those near the boundary of the Gribov region there will be a strong infrared enhancement
of the instantaneous interaction, which may cause confinement. This old observation has recently 
received support from several lattice computations\cite{CZ,GO}.

It is known that the non-Abelian Coulomb interaction is renormalization group invariant and that it
generates the complete running coupling of QCD\cite{Z2}. Furthermore, quarks decouple from
transverse gluons in the static limit and therefore this interaction must generate
confinement.  Indeed, Zwanziger has recently shown that the non-Abelian Coulomb interaction
provides an upper bound to the Wilson loop interaction\cite{Z3} and has conjectured that
this bound is saturated (numerical evidence in favour of this conjecture is provided in Ref.
\cite{CZ}).

It is clear that the instantaneous interaction is an important element of QCD. Recently an
analytical attempt to understand this operator has been made in the Greens function approach\cite{ss7}. The central idea was to compute the kernel appearing in Eq. \ref{coul} by summing all diagrams
which contribute at leading order in the infrared\cite{ss}. The resulting Dyson equation was
solved with the gluon propagator equation (which was obtained with the
aid of a Gaussian Ansatz for the pure gauge vacuum). Numerical solution of the coupled nonlinear
integral equations yielded a nontrivial gluonic quasiparticle dispersion relation and a linearly
confining instantaneous interaction in remarkable agreement with lattice Wilson loop results\cite{ss7}. These results are in agreement with the Kugo-Ojima confinement criteria, namely the inverse 
Faddeev-Popov operator (which is the Coulomb gauge analogue of the ghost propagator) is infrared
enhanced.  The infrared behaviour of the transverse gluon propagator is more problematic, but it
is now understood to arise from strong modifications of the vacuum Ansatz near the Gribov 
boundary\cite{s}.

\subsection{Vortices}

Singular gauge transformations generate local objects of great interest. For example, 
zero-dimensional gauge dislocations are associated with instantons, one-dimensional with
monopoles, and two-dimensional with vortices. A vortex is a sheet (in 4-d)
of infinite field strength which is associated with a $Z_{N_c}$ singular gauge transformation
(see Ref. \cite{KL0} for more details). It has been postulated that vortices drive confinement\cite{vortices} because they are localized field configurations which percolate the lattice.
The argument is quite general and relies on the fact that localized field distributions
contribute independently to the expectation value of the Wilson loop operator, and hence 
yield an area law interaction. 

There are many other attractive features of the vortex confinement picture, for example, deconfinement may be viewed as a vortex de-percolation phase transition\cite{KL2} and
the vortex density has been shown to scale in maximal centre gauge, thereby establishing their physical nature\cite{scaling}.
Lastly, it has recently been shown that monopoles may be understood in terms of vortex self intersections, thereby introducing topological field configurations into the discussion\cite{HR}.

\begin{figure}[h]
\includegraphics[height=5 true cm]{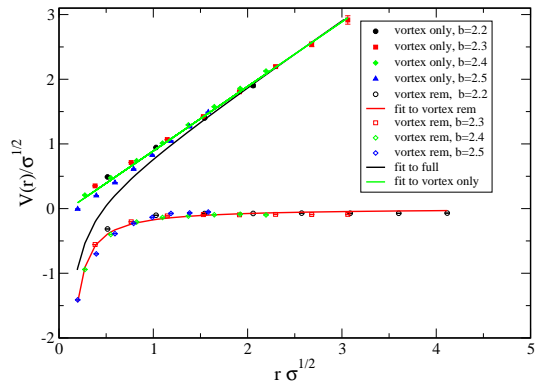}
\caption{The SU(2) Wilson Loop Potential with and without Vortices\cite{KL0}.}
\label{vortex}
\end{figure}

This compelling picture is strongly supported by lattice computations permitted by recent 
advances in lattice gauge fixing algorithms.  In particular figure \ref{vortex} shows that the 
Wilson loop string tension is maintained if one projects onto vortex gauge configurations
(upper green line)
{\it and} that linear confinement disappears if vortices are removed (lower red line). 

\section{Conclusions}

Great strides have been made recently in the effort to understand confinement. New lattice
algorithms allow a detailed study of the importance of different model gauge configurations
and of the onset of string-like behaviour in flux tubes. Multiquark states are also beginning to
be probed. At the same time, analytical methods are rapidly approaching the time when
quantitatively accurate computations of nonperturbative phenomena are possible. And we
are already seeing convincing demonstrations of the general features of confinement. 

Of course much remains to be done. The analytical formalisms need to examine the robustness
of the assumed truncations and the effects of topology and boundary conditions. At the same
time, these approaches can check and compare with a new breed of fixed-gauge lattice results.
There are many open issues  in the lattice approach a well. For example, too many models of
the QCD vacuum seem to be supported by the lattice and one needs to find some commonality
among the different mechanisms and gauges.  It is hoped and likely that issues such as
the nature of baryonic flux tubes, the infrared behaviour of the Landau gauge gluon propagator,
the onset of stringiness in flux tubes (is it at 1/2 or 4 fermi?), and the nature of the
non-Abelian Coulomb interaction will all be resolved in the short term.

\begin{theacknowledgments}

I am grateful to the organizers of Hadron03 for the invitation to speak on
such a fascinating topic and for providing a wonderful venue for discussing 
hadronic physics.  I wish to thank Reinhard Alkofer, Pierre van Baal, Jeff Greensite,
Hugo Reinhardt, Adam Szczepaniak, and Daniel Zwanziger for many illuminating discussions 
on this topic.
This work was supported by the DOE under contracts DE-FG02-00ER41135  and  DE-AC05-84ER40150.

\end{theacknowledgments}



\bibliography{sample}

\end{document}